\documentclass[12pt]{article}
\usepackage{amsmath}
\usepackage{graphicx}
\usepackage{amssymb} 
\newcommand{\Mat}[1]{{{\boldsymbol{#1}}}}
\newcommand{\abs}[1]{\left\vert#1\right\vert}
\def\be{\begin{equation}}
\def\ee{\end{equation}}
\def\dd{\mathrm{d}}
\title{{\bf Post-Newtonian equation for the energy levels of a Dirac particle in a static metric}}

\bigskip

\author{
Mayeul Arminjon\\
\small\it Dipartimento di Fisica, Universit\`a di Bari,\\
\small\it and Istituto Nazionale di Fisica Nucleare, Sezione di Bari,\\
\small\it Via Amendola 173, I-70126 Bari, Italy.
\footnote{
\ On leave from Laboratoire ``Sols, Solides, Structures''
(Unit\'e Mixte de Recherche of the CNRS),
 BP 53, F-38041 Grenoble cedex 9, France.
} 
}
\date{ }
\begin{document}

\maketitle


\begin{abstract}
\noindent We study first the Hamiltonian operator H corresponding to the Fock-Weyl extension of the Dirac equation to gravitation. When searching for stationary solutions to this equation, in a static metric, we show that just one invariant Hermitian product appears natural. In the case of a space-isotropic metric, H is Hermitian for that product. Then we investigate the asymptotic post-Newtonian approximation of the stationary Schr\"odinger equation associated with H, for a slow particle in a weak-field static metric. We rewrite the expanded equations as one equation for a two-component spinor field. This equation contains just the non-relativistic Schr\"odinger equation in the gravity potential, plus correction terms. Those ``correction" terms are of the same order in the small parameter as the ``main" terms, but are numerically negligible in the case of ultra-cold neutrons in the Earth's gravity.\\

\end{abstract}


\section {Introduction}

Quantum mechanics in a gravitational field is increasingly becoming an experimentally relevant subject \cite{KieferWeber2005}. Due to the weakness of the gravity interaction, the effects of gravity are more clearly seen in experiments with neutral particles: neutrons, and also atoms or molecules. The quantum-mechanical phase shift of neutrons due to their interaction with the Earth's gravitational field has been measured since a long time, thanks to interferometric experiments \cite{COW1975}. \{Similar effects, due to a non-inertial (rotational) motion or to the Earth's gravity acceleration, have been observed in atom interferometry \cite{RiehleBorde1991,KasevichChu1991}.\} More recently, the quantization of the energy levels of ultra-cold neutrons in this same terrestrial gravity field has been revealed by measuring their transmission through a horizontal slit \cite{Nesvizhevsky2002,Nesvizhevsky2003}. New, more precise experiments are being foreseen in order to accurately determine the energy levels \cite{Nesvizhevsky2005}. Up to now, the theoretical analyses of these experiments have been done in the framework of the non-relativistic Schr\"odinger equation in the Newtonian gravity potential \cite{Nesvizhevsky2002,OverhauserColella1974,LuschikovFrank1978,WestphalAbele2006}. This is justified by the smallness of the velocity of the neutrons used in these experiments (5 m/s to 10 m/s in the transmission measurements \cite{Nesvizhevsky2002,Nesvizhevsky2003}, ca. $3\times 10^2\ $m/s in the interferometric experiment \cite{COW1975}), and by the weakness of the gravitational field of the Earth.\\

However, gravity is currently described by relativistic theories like general relativity (GR). In the framework of GR and other theories based on Einstein's equivalence principle, the wave equations of relativistic quantum mechanics, i.e. essentially the Klein-Gordon equation and the Dirac equation, are adapted to gravitation by rewriting them in a generally-covariant way. For the Klein-Gordon equation, this ``covariantization" is ambiguous \cite{BirrellDavies}, due to the fact that covariant derivatives do not commute. In the case of the Dirac equation with the spinor transformation, this procedure leads to the equation independently proposed by Fock \cite{Fock1929b} and by Weyl \cite{Weyl1929b}, hereafter named the Dirac-Fock-Weyl (DFW) equation. On the other hand, an alternative gravitational Dirac equation has been recently proposed (for the case of a static metric) \cite{A37}. It is based on a direct derivation from the principles of wave mechanics, instead of using the equivalence principle. The new equation definitely violates the latter principle, for it does not reduce to the flat-space-time Dirac equation in a local ``freely falling" frame \cite{A37}. \\

Since neutrons are particles with spin 1/2, one may hope that, at least in the absence of an external electromagnetic field, their behaviour in the gravitational field should be correctly described by one of these two gravitational Dirac equations---and, of course, it would be interesting to know which one of the two. The aim of this paper is to derive, for the standard, DFW equation, a Schr\"odinger-type equation which will allow to compute the stationary energy levels. (The corresponding equation for the alternative equation \cite{A37} will be derived in a forthcoming work.) The equation to be derived exhibits the first correction terms with respect to the non-relativistic stationary Schr\"odinger equation in the Newtonian gravity potential. This does not appear to have been done before, although some amount of work has been devoted to studying the weak-field and/or non-relativistic limit of the DFW equation (see, among others, Refs. \cite{deOliveiraTiomno1962,VarjuRyder1998,Obukhov2001,SilenkoTeryaev2005}). These works did result in proposals for an approximate Hamiltonian operator, hence one should be able to find easily the approximate equation for the stationary energy levels by using those works. However, the approximation scheme was not made explicit in these works, so that, in the approximate equations, one cannot easily assign orders to the different terms, with respect to a small parameter---as is necessary to ensure that the first corrections with respect to the non-relativistic Schr\"odinger equation are consistently evaluated. In addition, these works were concerned with the Hamiltonian for the four-dimensional complex wave function, whereas we will verify in subsect. \ref{correct-adjoint} that (not unexpectedly) the solutions of the stationary DFW equation have only two independent complex components. Finally, each of the Hamiltonians which were used in Refs. \cite{deOliveiraTiomno1962,VarjuRyder1998,Obukhov2001,SilenkoTeryaev2005}, had been obtained from the starting Hamiltonian of the DFW equation by using a non-unitary transformation \cite{deOliveiraTiomno1962,VarjuRyder1998,Obukhov2001}. In the absence of another interpretation, this procedure might be seen as coordinate-dependent \cite{A38v1}.\\

The paper begins (Sect. \ref{HermitianHamiltonian}) with a discussion of the Hamiltonian operator for the DFW equation. After noting the invariance of this operator, we investigate the question of the relevant scalar product, in the context of the search for stationary energy levels in a static metric. The link with the procedure adopted in Refs. \cite{deOliveiraTiomno1962,VarjuRyder1998,Obukhov2001} is made through the recent interpretation of this procedure by Leclerc \cite{Leclerc2006a}. In Sect. \ref{StationaryPNA}, the weak-field and non-relativistic limit of the stationary DFW equation in a static metric is being studied, i.e., a slow Dirac particle is considered, in a static metric that differs little from a flat metric. The application to ultra-cold neutrons in the Earth's gravity is discussed. Our conclusion makes Sect. \ref{Conclusion}.

\section{Hermitian Hamiltonian for the Dirac-Fock-Weyl equation} \label{HermitianHamiltonian}
\subsection{Starting Hamiltonian for the DFW equation}
The DFW equation may be written as \cite{deOliveiraTiomno1962,BrillWheeler1957+Corr,VillalbaGreiner2001}:
\be \label{DFW}
\left ( i\gamma^\mu D_\mu -M \right) \psi =0, \qquad D_\mu \equiv \partial _\mu -\Gamma _\mu, \quad M \equiv mc/\hbar,
\ee
where $m$ is the rest mass of the quantum particle, and where the ``deformed" Dirac matrices $\gamma ^\mu$ satisfy the anticommutation relation
\be \label{Clifford}
\gamma ^\mu \gamma ^\nu + \gamma ^\nu \gamma ^\mu = 2g^{\mu \nu}\,{\bf 1}_4, \qquad {\bf 1}_4 \equiv \mathrm{diag}(1,1,1,1),
\ee
with $(g^{\mu \nu})\equiv (g_{\mu \nu})^{-1},\quad \Mat{g}=(g_{\mu \nu})$ being the (curved) space-time metric, and where the $\Gamma _\mu$ matrices are given by
\be \label{Gamma_mu}
\Gamma _\mu = \frac{1}{4} g_{\lambda \alpha } b^\beta _{\nu ,\mu} a^\alpha _\beta \, s^{\lambda \nu} - \frac{1}{4} \Gamma _{\lambda \nu \mu}\, s^{\lambda \nu}, 
\ee
in which the $\Gamma _{\lambda \nu \mu}$ 's are the first-kind Christoffel symbols, $ s^{\lambda \nu} \equiv \frac{1}{2} \left (\gamma ^\lambda \gamma ^\nu - \gamma ^\nu \gamma ^\lambda \right )$, and the matrices $A=(a^\alpha _\beta)$ and $B=(b^\beta _\nu)$ (with $a^\alpha _\beta b^\beta _\nu = \delta ^\alpha _\nu $) transform the natural basis $e_\alpha = \frac{\partial }{\partial x^\alpha }$ into the local ``tetrad" $u_\beta $ and conversely: $u_\beta = a^\alpha _\beta e_\alpha $, $\ e_\alpha = b^\beta _\alpha u_\beta $, the tetrad being orthonormal: 
\footnote{\ 
One may go from the $(1,-1,-1,-1)$ signature, used here and in Refs. \cite{deOliveiraTiomno1962,VarjuRyder1998,Obukhov2001,SilenkoTeryaev2005}, to the $(-1,1,1,1)$ signature, used in Refs. \cite{BrillWheeler1957+Corr,VillalbaGreiner2001}, by setting: $\Mat{\eta '} = -\Mat{\eta }$, $\Mat{g'} = -\Mat{g}$, and $ \gamma '^\mu =-i\gamma ^\mu$. The bases and the matrices $A,B$ stay unchanged, as well as Eqs. (\ref{Clifford})-(\ref{tetrad})$_1$ (of course with primes), while Eq. (\ref{DFW}) takes the form assumed in Refs. \cite{BrillWheeler1957+Corr,VillalbaGreiner2001}.
}
\be \label{tetrad}
\Mat{g}(u_\beta , u_\nu) = a^\alpha _\beta a^\mu _\nu g_{\alpha \mu} = \eta _{\beta \nu}, \qquad \Mat{\eta} \equiv \mathrm{diag}(1,-1,-1,-1). 
\ee
 The ``deformed" Dirac matrices are related to the ``flat" ones $\tilde{\gamma }^\beta $, that satisfy Eq. (\ref{Clifford}) with $\Mat{g}=\Mat{\eta}$, by 
\be \label{flat-deformed}
\gamma ^\alpha  = a^\alpha _\beta \ \tilde{\gamma }^\beta.
\ee

\vspace{5mm}
Multiplying (\ref{DFW}) by $\gamma ^0$ on the left and using (\ref{Clifford}), one gets the DFW equation in Schr\"odinger form:
\be \label{Schrodinger-DFW}
i\hbar \frac{\partial \psi }{\partial t}= \mathrm{H}\psi,
\ee
with 
\be \label{Hamilton-DFW}
 \mathrm{H} \equiv  \frac{mc^2}{g^{00}}\gamma ^0 -i\frac{\hbar c}{g^{00}}\gamma ^0\gamma ^j D_j +i\hbar c\Gamma _0 .
\ee
(Spatial indices are denoted by Latin letters, space-time indices by Greek ones; and $x^0=ct$.) The Hamiltonian operator (\ref{Hamilton-DFW}) makes sense whether Eq. (\ref{DFW}) is satisfied or not, and may be rewritten as 
\be \label{Hamilton-DFW-space-covariant}
\mathrm{H} = i\hbar \partial_ t  + \frac{mc^2}{g^{00}}\gamma ^0 -i\frac{\hbar c}{g^{00}}\gamma ^0\gamma ^\mu  D_\mu.   
\ee
The DFW equation (\ref{DFW}) is covariant under a general coordinate change: $x'^\mu = \phi ^\mu ((x^\nu))$, under which the spinor $\psi $ is left {\it invariant} \cite{deOliveiraTiomno1962,BrillWheeler1957+Corr}, while the $\gamma ^\mu $ matrices transform like a vector, i.e., $\gamma'^{\mu} =\gamma ^\nu \partial x'^\mu /\partial x^\nu $ \cite{BrillWheeler1957+Corr}. (This is because the tetrad is left unchanged in a coordinate change.) Hence, $\gamma ^\mu  D_\mu \psi $ should also be invariant. It is in fact so: a tedious calculation allows one indeed to check that $\Gamma _\mu$ [Eq. (\ref{Gamma_mu})], hence also $D_\mu \psi $, transforms like a covector. It follows then from (\ref{Hamilton-DFW-space-covariant}) that the modified spinor field after applying the Hamiltonian operator: $\mathrm{H}\psi $, remains also {\it invariant} under purely {\it spatial} coordinate changes, $x'^j = \phi ^j ((x^k)), \ x'^0=x^0$, or ${\bf x}'={\bf F}({\bf x})$, $t'=t$. That is, 
\be \label{Hamiltonian-invariance}
(\mathrm{H}'\psi')(t,{\bf x}') =(\mathrm{H}\psi)(t,{\bf x}), \qquad [{\bf x}'\equiv {\bf F}({\bf x})],
\ee
with in fact $\psi '(t,{\bf x}')=\psi(t,{\bf x})$.\\

When the metric is diagonal: $\Mat{g}=\mathrm{diag}(a_\mu)$, one may define the tetrad from the matrix \cite{VillalbaGreiner2001}
\be \label{tetrad-g-diagonal}
A = \mathrm{diag}(1/\sqrt{ \abs{a_\mu} } ).
\ee
Using this and (\ref{flat-deformed}), we find that the first sum in Eq. (\ref{Gamma_mu}) vanishes, and that
\be \label{Gamma_mu-g-diagonal}
\Gamma _\mu = -\frac{1}{4} \frac{a_{\mu ,\nu}}{ \sqrt{\abs{a_\mu a_\nu}} } \tilde{s}^{\mu \nu}\ (\mathrm{no\ sum\ on\ }\mu), \qquad \tilde{s}^{\mu \nu} \equiv \frac{1}{2} \left (\tilde{\gamma} ^\mu \tilde{\gamma} ^\nu - \tilde{\gamma}^\nu \tilde{\gamma} ^\mu \right ).
\ee
Let us specialize the Hamiltonian (\ref{Hamilton-DFW}) to the particular diagonal metric assumed by Obukhov \cite{Obukhov2001}:
\be \label{g-Obukhov}
a_0 = [V({\bf x})]^2, \qquad a_j =-[W({\bf x})]^2, \qquad {\bf x} \equiv  (x^j) \quad (j=1,2,3).
\ee
This is a static metric, i.e., it verifies
\be \label{static-metric}
g_{\mu \nu}=g_{\mu \nu}({\bf x}) \qquad \mathrm{and}\qquad g_{0j}=0, 
\ee
whence $g^{00}=g_{00}^{-1}$, thus $g^{00}=V^{-2}$ for the metric (\ref{g-Obukhov}). We find from (\ref{Gamma_mu-g-diagonal}):
\be \label{Gamma_mu-g-Obukhov-0}
\Gamma _0 = -\frac{V_{,j}}{2W}\alpha ^j, \qquad \alpha ^j \equiv \tilde{\gamma}^0 \tilde{\gamma}^j,
\ee
\begin{equation} \label{Gamma_mu-g-Obukhov-j}
\Gamma _j = \frac{W_{,k}}{2W} \tilde{s}^{j k}=-i\varepsilon _{j k l}\frac{W_{,k}}{2W}\Sigma ^l, \qquad \Sigma ^l \equiv \left (\begin{array}{cc} \sigma ^l \ 0 \\ 0\ \sigma ^l \end{array} \right),
\end{equation}
where $\sigma ^l (l=1,2,3)$ are the Pauli matrices. (Beginning with (\ref{Gamma_mu-g-Obukhov-j}), we shall use the standard set \cite{Obukhov2001,BjorkenDrell1964} of the matrices $\tilde{\gamma}^\mu$.) Putting (\ref{Gamma_mu-g-Obukhov-0}) and (\ref{Gamma_mu-g-Obukhov-j}) in the Hamiltonian (\ref{Hamilton-DFW}), and using again the diagonal tetrad (\ref{tetrad-g-diagonal}) with (\ref{g-Obukhov}) in the definition (\ref{flat-deformed}) of the deformed Dirac matrices, we get
\be \label{Hamilton-Obukhov-1}
\mathrm{H} = mc^2\beta V -i\hbar c \alpha ^j \left ( \frac{V_{,j}}{2W} + \frac{V}{W}\partial _j +i \frac{V}{2W^2} \varepsilon _{j k l}W_{,k} \Sigma ^l\right ), \quad \beta \equiv \tilde{\gamma }^0.  
\ee
Using then the standard definitions of $\alpha ^j$ and $\Sigma ^l$, one checks easily that Obukhov's starting Hamiltonian \{\cite{Obukhov2001}, Eq. (12)\} is recovered:
\be \label{Hamilton-Obukhov-2}
\mathrm{H} = mc^2\beta V -i\hbar c \alpha ^j \left ( \frac{V_{,j}}{2W} +\frac{V}{W^2} W_{,j} \right )-i\hbar c \alpha ^j  \frac{V}{W}\partial _j .  
\ee 

\subsection{Stationary energy levels in a static metric} \label{energy-levels}

For a static metric, the time coordinate $t$ in which we have (\ref{static-metric}) is {\it unique} (up to a scale factor) \cite{A16}. Thus, if we postulate a stationary wave:
\be \label{stationary-wave}
\psi (t,{\bf x}) = \phi (t)\,a({\bf x}),
\ee
this decomposition (\ref{stationary-wave}) must use that ``static time" $t$. \{For a time-independent but non-static metric, in contrast, the decomposition (\ref{stationary-wave}) would not be well-defined, since it might use any of the different time coordinates compatible with the time-independence \cite{L&L}, say $t$ or $t'=t+f ({\bf x})$.\} Moreover, for any time-independent metric, the Hamiltonian operator (\ref{Hamilton-DFW}) does not contain time derivatives, and its coefficients do not depend on time. Hence, as is well-known, substituting the stationary wave (\ref{stationary-wave}) into the Schr\"odinger-type equation (\ref{Schrodinger-DFW}), leads to 
\be \label{time-dependence}
\phi  (t)=C\exp(-iEt/\hbar),
\ee
while the amplitude function $a$ \{which may well have several components, as is the case here, provided that the time-dependence $\phi (t)$ is a scalar \cite{B15}\} is a solution of the eigenvalue problem
\be \label{eigenvalue-E}
\mathrm{H}a = Ea \equiv \hbar \omega a.
\ee
Equations (\ref{stationary-wave})-(\ref{eigenvalue-E}) define the stationary energy levels associated with the Hamiltonian $\mathrm{H}$---provided, of course, that the ``energy levels" themselves, i.e., the eigenvalues $E$ of problem (\ref{eigenvalue-E}), are real: if that is not the case, the time-dependence (\ref{time-dependence}) contains a (real) exponential term, hence the solutions (\ref{stationary-wave}) can in no way be qualified stationary. Hermitian operators make the only well-known class of operators for which we can a priori state  (thus before calculating the spectrum) that the eigenvalues are real.
\footnote{\ 
But, of course, there are plenty of non-Hermitian operators with all real eigenvalues---although, within the important class of ``normal" operators (i.e., those which commute with their adjoint), an operator is Hermitian if and only if its spectrum is real \cite{DieudonneTome2}.
}
\subsection{Scalar product}

The Hermitian conjugate of an operator depends, of course, on the (Hermitian) scalar product which is used on the domain of this operator. Thus arises the question of the scalar product $( a; b )$ with respect to which $\mathrm{H}$ may, or may not, be a Hermitian  operator. That scalar product should not involve derivatives. Otherwise, when calculating $(\mathrm{H}a ;b )$, higher-order derivatives would arise. Hence, it has necessarily the following general form:
\be \label{Hermitian-general}
(a ; b ) \equiv \int_\mathrm{M} (a (x). b (x)) \ \dd {\sf V}(x),
\ee
where $(y.z)$ is a Hermitian product defined for arrays $y$ and $z$ of four complex numbers, and where ${\sf V}$ is some volume measure defined on the ``space" manifold M, i.e., on the preferred spatial section defined by the static metric. The scalar product $(y.z)$ used in the integration (\ref{Hermitian-general}) might a priori depend on the integration variable, thus on the point $x \in \mathrm{M}$, say $(y.z)_{ x}$. However, in the case of the DFW equation, the value $\psi (t,{\bf x})$ of the spinor is invariant under general coordinate changes [hence the amplitude $a({\bf x})=a(x)$ of a stationary $\psi $ (\ref{stationary-wave}) is invariant under spatial coordinate changes]. As is usually phrased, this means that it belongs to some ``internal space" independent of $ x$---say, simply to ${\sf C}^4$. Hence, in the present case, the product $(y.z)$ should {\it not} depend on $x$. The only natural product is then the canonical scalar product on ${\sf C}^4$,
\be \label{C4-product}
(y.z) \equiv y^{\mu *} z^{\mu} \equiv y^\dagger  z.
\ee
In contrast, if $\psi $ were a 4-vector under coordinate transforms, as is in fact the case for the alternative gravitational Dirac equation \cite{A37}, the use of (\ref{C4-product}) would be incompatible with the {\it coordinate-independence} of the product (\ref{Hermitian-general}). The coordinate-independence of (\ref{Hermitian-general}) demands also that the volume measure ${\sf V}$ be invariant under the allowed coordinate changes, i.e., under the purely spatial coordinate changes (which are those compatible with the static character of the metric). But, for a general static metric, there is only one such invariant volume measure, whose expression in a given coordinate system is
\be \label{invariant-V}
\dd {\sf V}({\bf  x}) = \sqrt{h({\bf x})}\ \dd ^3{\bf x}, \quad h \equiv \mathrm{det}(h_{jk}),
\ee
where $(h_{jk})$ is the component matrix of the metric $\Mat{h}$ on M associated with the static space-time metric $\Mat{g}$ (with components $g_{\mu \nu}$), i.e., $h_{jk}=-g_{jk}$ in static-compatible coordinates. Thus, we find that there is only one coordinate-independent scalar product that appears naturally when investigating the Hermitian character of the Hamiltonian associated with the DFW equation for a static metric---namely, the product (\ref{Hermitian-general}), specified by the definitions (\ref{C4-product}) and (\ref{invariant-V}). 
\footnote{\
Tho ensure that (\ref{Hermitian-general}) is well-defined, one must assume square-integrable functions: $a$ and $b  \in \mathrm{L}^2(\mathrm{M},{\sf V})$.
}
\subsection{Link with the literature} \label{link-literature}
In Refs. \cite{deOliveiraTiomno1962,AudretschSchaefer1978,VarjuRyder1998,Obukhov2001}, a product with the same form (\ref{Hermitian-general}), and with (implicitly) the same definition (\ref{C4-product}) for the ``local" product, is being used. Each of these works considers a particular coordinate system, in which the space-time metric is assumed to have a special form---e.g., the form (\ref{g-Obukhov}) in Ref. \cite{Obukhov2001}. Each in their specific coordinate system, these authors use the coordinate volume measure $\dd ^3{\bf x}$, instead  of the invariant volume measure (\ref{invariant-V}). Thus, they use the scalar product
\be \label{Hermitian-coordinate}
(a \mid b ) \equiv \int_{{\sf R}^3} (a ({\bf x}). b ({\bf x})) \ \dd ^3{\bf x}.
\ee 
If one interprets this as the basic, starting, scalar product used by these authors, he notes that it is coordinate-dependent, and that the Hamiltonian (\ref{Hamilton-DFW}) is, in general, not Hermitian for this product---e.g., the Hamiltonian (\ref{Hamilton-Obukhov-2}) is not Hermitian for the product (\ref{Hermitian-coordinate}), if the metric has the form (\ref{g-Obukhov}). Then, the non-unitary transform used by these authors:
\be \label{Non-unitary-transform}
\breve{\psi} = T\psi, \qquad \breve{\mathrm{H}} = T\mathrm{H}T^{-1},
\ee
with
\be \label{T-non-unitary}
T = \left(\frac{-g}{g_{00}}\right)^{1/4}, \qquad g \equiv \mathrm{det}(g_{\mu \nu })
\ee
[and, in particular,
\be \label{A-Obukhov}
T=W^{3/2}
\ee
for the metric (\ref{g-Obukhov})], is interpreted as a modification of the Hamiltonian with the aim of getting a Hermitian Hamiltonian \cite{A38v1}. This interpretation leads to a problem: the modified Hamiltonian $ \breve{\mathrm{H}}$ defined by Eq. (\ref{Non-unitary-transform}) is, in general, not Hermitian any more if one transports it to another coordinate system and uses the scalar product defined by the same Eq. (\ref{Hermitian-coordinate}) taken in the new coordinates \cite{A38v1}. \\

However, a more relevant interpretation of the works \cite{deOliveiraTiomno1962,AudretschSchaefer1978,VarjuRyder1998,Obukhov2001} has been suggested recently by Leclerc \cite{Leclerc2006a} (Sect. 3 in that reference). He notes that the transformation (\ref{Non-unitary-transform})-(\ref{T-non-unitary}) gets the scalar product
\be \label{Hermitian-gamma}
(a\parallel b) \equiv \int_{{\sf R}^3} \sqrt{-g}\ a ({\bf x})^\dagger \tilde{\gamma }^0\gamma ^0 b ({\bf x}) \ \dd ^3{\bf x}
\ee
to the scalar product (\ref{Hermitian-coordinate}), in an isometric way, i.e.
\be \label{scalar-conserved}
 (\breve{a}\mid \breve{b}) = (a\parallel b).
\ee
This is indeed easy to check in the case of a diagonal tetrad (\ref{tetrad-g-diagonal}) (implying a diagonal metric), for we get then from (\ref{Clifford}) and (\ref{T-non-unitary}):
\be \label{gamma0tilde-gamma0-diagonal}
\sqrt{-g}\, \tilde{\gamma }^0\gamma ^0 = T^2 \,{\bf 1}_4. 
\ee
Due to (\ref{scalar-conserved}) and (\ref{Non-unitary-transform}), we have also
\be \label{scalar-H-conserved}
 (\breve{\mathrm{H}}\breve{a}\mid \breve{b}) = (\mathrm{H}a\parallel b).
\ee
Thus, if the operator $\mathrm{H}$ is Hermitian for the product (\ref{Hermitian-gamma}), then so is also $\breve{\mathrm{H}}$ for the product (\ref{Hermitian-coordinate}) \cite{Leclerc2006a}, and conversely; in addition, the two operators have the same spectrum. As noted by Leclerc \cite{Leclerc2006a}, ``the
hermiticity properties of the operators can be directly read off" with the flat product (\ref{Hermitian-coordinate}): for example, see Eq. (14) in Ref. \cite{Obukhov2001}. The scalar product (\ref{Hermitian-gamma}) arises from the conservation law associated with the standard DFW current, and it may be rewritten in a coordinate-independent way \cite{Leclerc2006a} (for a given hypersurface, defined in the starting coordinate system by $x^0=\mathrm{Const}.$). Here, we show that the product (\ref{Hermitian-gamma}) coincides, in the case of a diagonal tetrad (\ref{tetrad-g-diagonal}), with the coordinate-independent product (\ref{Hermitian-general})-(\ref{invariant-V}) that we introduced. We have in the most general case \cite{L&L}:
\be
T^4\equiv \frac{-g}{g_{00}}=h.
\ee
For a diagonal tetrad, we have Eq. (\ref{gamma0tilde-gamma0-diagonal}), whence indeed from (\ref{invariant-V}):
\be \label{prodMA=prodML}
(a\parallel b) = (a; b).
\ee
We conclude that, in the case of a static metric which can be reduced to the diagonal form (which is the case considered here and in Refs. \cite{VarjuRyder1998,Obukhov2001}), the invariant scalar product (\ref{Hermitian-general})-(\ref{invariant-V}) can be rewritten in the form (\ref{Hermitian-gamma}). Then, the non-unitary transform (\ref{Non-unitary-transform}) is indeed ``a purely mathematical operation which does not affect the physics," and which transforms that product to the flat product (\ref{Hermitian-coordinate}), in terms of which the hermiticity of the Hamiltonian operator is easily checked \cite{Leclerc2006a}. In particular, in the case of the metric (\ref{g-Obukhov}), this transformation brings the Hamiltonian (\ref{Hamilton-Obukhov-2}) to an explicitly-Hermitian form \cite{Obukhov2001}, hence we know that the starting Hamiltonian (\ref{Hamilton-Obukhov-2}) is Hermitian for the invariant product (\ref{Hermitian-general})-(\ref{invariant-V}). Nevertheless, it is interesting to show that the adjoint of the Hamiltonian (\ref{Hamilton-Obukhov-2}) for that product can be determined directly.
\footnote{\ 
When investigating the hermiticity of the Dirac(-Fock-Weyl) Hamiltonian, or, more exactly, when checking whether $ (\mathrm{H}\psi \parallel \psi ) = (\psi \parallel \mathrm{H}\psi)$, Leclerc \cite{Leclerc2006a} assumes that $\psi $ verifies the wave equation. However, this identity [and in fact the more general identity $ (\mathrm{H}\psi \parallel \varphi  ) = (\psi \parallel \mathrm{H}\varphi )$] has to be verified by general functions $\psi $ (and $\varphi $) in the domain of $\mathrm{H}$, not merely by ones satisfying the wave equation. Moreover, in the context of the search for stationary solutions (\ref{stationary-wave}), the relevant hermiticity applies to {\it spatial} wave functions $a$. The associated time-dependent wave function (\ref{stationary-wave}) obeys the wave equation iff $a$ is an eigenfunction of $\mathrm{H}$, hence checking the hermiticity for such wave functions is not enough.
}

\subsection{Adjoint of the Hamiltonian (direct method)} \label{correct-adjoint}
Thus, let us compute the adjoint $\mathrm{H}^\dagger$ of the operator $\mathrm{H}$, for the scalar product (\ref{Hermitian-general})-(\ref{invariant-V}). To do that, we need to know the adjoint operator of the partial derivative $\partial _j$ (or, equivalently, that of the momentum operator $p_j\equiv -i\hbar \partial _j$). Of course, the operator $\partial _j$ is coordinate-dependent, and so will be its adjoint. But, since the Hamiltonian (\ref{Hamilton-DFW}) verifies Eq. (\ref{Hamiltonian-invariance}), its application to a merely space-dependent spinor $a({\bf x})$ verifies
\be \label{H-Invariance-space} 
\mathrm{H}'a' ({\bf x}') = \mathrm{H}a ({\bf x}).
\ee 
The scalar product (\ref{Hermitian-general})-(\ref{invariant-V}) being invariant too, the resulting adjoint operator $\mathrm{H}^\dagger$ is also invariant. I.e., if one makes an admissible (thus purely spatial) coordinate change, ${\bf x}'={\bf F}({\bf x})$, $t'=t$, then the adjoint operators will correspond by Eq. (\ref{H-Invariance-space}) (with $\mathrm{H}^\dagger$ in the place of $\mathrm{H}$). We determine first the adjoint $\partial _j^\ddagger $ of the operator $\partial _j$, considered as acting on {\it scalar} functions, thus with respect to the invariant scalar product that corresponds to (\ref{Hermitian-general}) for scalar functions,
\be \label{Hermitian-general-scalar}
(a , b ) \equiv \int_\mathrm{M} a ^*\,b  \ \dd {\sf V}.
\ee
If the scalar function $f$ cancels outside a bounded domain, or more generally if [in the coordinates $(x^j)$], $f({\bf x})\sqrt{h({\bf x})}=o(1/r^2)$ as $r\equiv \abs{ {\bf x}} \rightarrow \infty $,
\footnote{\
Thus, we assume that the space M is diffeomorphic to ${\sf R}^3$, and that the chart $x \mapsto {\bf x}= (x^j)$ covers M. 
}
then one has
\be \label{divergence-theorem-curved}
\int_\mathrm{M} \frac{1}{\sqrt{h}}(f\,\sqrt{h})_{,j} \ \dd {\sf V} =\int_{{\sf R}^3} \frac{\partial }{\partial x^j}\left [f({\bf x})\sqrt{h({\bf x})} \right] \ \dd ^3 {\bf x}=0.
\ee
(This is got by applying the divergence theorem to the vector field $ f({\bf x})\sqrt{h({\bf x})}\ {\bf e}_j$ with the ball $\mathrm{B}({\bf 0}, r)$ in the Euclidean space ${\sf R}^3$.) By using this result with $f=a^* b $, we obtain
\be \label{div-a*b}
\int_\mathrm{M} \left( a^* _{,j}b +a^* b _{,j} \right ) \dd {\sf V} + \int_\mathrm{M} a^* b  \frac{h_{,j}}{2h} \ \dd {\sf V}=0.
\ee
As we know, the adjoint of an operator A [here one acting on scalar functions, thus with the scalar product (\ref{Hermitian-general-scalar})] is defined to be the operator $\mathrm{A}^\ddagger $ such that
\be \label{adjoint-definition}
(\forall a)\ (\forall b ) \quad (a ,\mathrm{A}b )=(\mathrm{A}^\ddagger a ,b ).
\ee
Hence, it follows from Eq. (\ref{div-a*b}) that
\footnote{\ 
Without going into detailed considerations on the domains of the operators, it is clear that the functions to which (\ref{div-a*b}) applies will be dense in the relevant domain.
}
\be \label{adjoint-partial-j}
\partial _j^\ddagger = -\partial _j -\frac{h_{,j}}{2h},
\ee
with, for the metric (\ref{g-Obukhov}),
\be \label{h,j/2h}
\frac{h_{,j}}{2h}=3\frac{W_{,j}}{W}.
\ee
It is easy to check that, when an operator A is extended from scalar functions to ``vector" ones by $\mathrm{A}.(a ^\mu)\equiv (\mathrm{A}a ^\mu)$, its adjoint $\mathrm{A}^\dagger $ for the product (\ref{Hermitian-general})-(\ref{invariant-V}) is obtained by extending in the same way the adjoint $\mathrm{A}^\ddagger $  for the product (\ref{Hermitian-general-scalar}). Also as expected, the adjoint, for the product (\ref{Hermitian-general})-(\ref{invariant-V}), of the operator defined by a complex matrix $M$, is the operator defined by the adjoint matrix, $M^\dagger \equiv (M^*)^\mathrm{T}$. Thus, in Eq. (\ref{Hamilton-Obukhov-2}), the first operator is Hermitian, the second one is anti-Hermitian, and the adjoint of the third one is [Eqs. (\ref{adjoint-partial-j})-(\ref{h,j/2h})]:
\footnote{\ 
The usual convention is followed for product operators: thus, the product operator
$\partial_j F$ transforms $a$ to $\partial_j(Fa) = (F\partial_j + F_{,j})a$, with $F_{,j}$ the mere multiplication by the
function $\partial F/
\partial x^j$.
}
\be 
\left [F \alpha ^j  (-i\partial_j) \right ]^\dagger = \left (-i \partial_j-3i \frac{W_{,j}}{W} \right )\alpha ^j F = \alpha ^j \left [-i (F\partial_j+F_{,j})-3i \frac{W_{,j}}{W}F \right ].
\ee
We find then easily that
\begin{eqnarray} \label{Hamilton-Obukhov-adjoint}
\mathrm{H}^\dagger  & = & mc^2\beta V -i\hbar c \alpha ^j \left ( F\partial_j+F_{,j}  +2\frac{W_{,j}}{W}F - \frac{V_{,j}}{2W}  \right )\\
& = & mc^2\beta V -i\hbar c \alpha ^j \left (\frac{VW_{,j}}{W^2} +\frac{V_{,j}}{2W}+ \frac{V}{W}\partial _j  \right )= \mathrm{H}.
\end {eqnarray}
Thus, we checked directly that the Hamiltonian operator (\ref{Hamilton-Obukhov-2}) is Hermitian for the invariant scalar product (\ref{Hermitian-general})-(\ref{invariant-V}). Therefore, to compute the stationary energy levels of the Dirac particle, we may consider the eigenvalue problem (\ref{eigenvalue-E}) with this starting Hamiltonian (\ref{Hamilton-Obukhov-2}). Using the standard set \cite{Obukhov2001,BjorkenDrell1964} of the matrices $\alpha ^j$ and $\beta $, and writing the 4-component amplitude function $a({\bf x})$ as a couple of 2-component functions, 
\be \label{two 2-spinors}
a({\bf x})=(\varphi ({\bf x}),\chi ({\bf x})), 
\ee
we may write explicitly (\ref{eigenvalue-E}) (after division by $\hbar c$) as
\be \label{eigenvalue-E-phi}
-i\frac{V}{W} \sigma ^j \chi _{,j}-i \left(\frac{V_{,j}}{2W}+\frac{VW_{,j}}{W^2} \right)\sigma ^j \chi = \left(\frac{\omega }{c}-MV \right)\varphi ,
\ee
\be \label{eigenvalue-E-chi}
-i\frac{V}{W} \sigma ^j \varphi  _{,j}-i \left(\frac{V_{,j}}{2W}+\frac{VW_{,j}}{W^2} \right)\sigma ^j \varphi  = \left(\frac{\omega }{c}+MV \right)\chi.
\ee 
Thus, $\chi $ is expressed as a function of $\varphi $ by Eq. (\ref{eigenvalue-E-chi}), and is eliminated by reporting this in (\ref{eigenvalue-E-phi}). In this sense, the stationary DFW equation also has only two independent complex degrees of freedom, as is well-known for the flat-space-time Dirac equation. 

\section{Post-Newtonian approximation (PNA) for stationary energy levels} \label{StationaryPNA}
\subsection{Framework: asymptotic PNA}
A PNA was already considered in the quantum domain by Kiefer \& Singh \cite{KieferSingh1991}, who expanded in powers of $c^{-2}$ the phase function in the ``minimally-coupled" gravitational Klein-Gordon equation, in a non-stationary situation. This approach was further developed by L\"ammerzahl \cite{Laemmerzahl1995}, who applied it to explore the modification of the coupling of matter to the electromagnetic field, which is induced by the presence of a (weak) gravitational field. Here, we are using the so-called ``asymptotic" PNA \cite{FutaSchutz,A36,A35}, i.e., we consider a {\it (conceptual) family} (S$_\lambda $) of systems. The aim of this approach is to give a mathematically clearer meaning to the postulated expansions, interpreting them indeed as asymptotic expansions. In the present case, each system S$_\lambda $ is constituted by a massive body producing the static metric, and by a Dirac particle which is in a stationary state in this metric. Since we are studying a ``test quantum-particle", assumed to not affect the metric, we need to know only the metric produced by the body: let $\Mat{g}^{(\lambda )}$ be the metric in system S$_\lambda $. We do not need to specify which theory of gravitation is assumed. We suppose that there is a coordinate system $t,{\bf x}$, such that, for any $\lambda $, the metric $\Mat{g}^{(\lambda )}$ has the form (\ref{g-Obukhov}) with the following (post-Newtonian) asymptotic expansion as $\lambda \rightarrow 0$:
\be \label{V-expansion}
V^{(\lambda )}(\mathbf{x}) \equiv  \left[g^{(\lambda )}_{00}\right]^{1/2}=1-\lambda\, \frac{U(\mathbf{x})}{c^2} +O(\lambda ^2),
\ee
\be \label{W-expansion}
W^{(\lambda )}(\mathbf{x}) \equiv  \left[-g^{(\lambda )}_{jj}\right]^{1/2}= 1+\lambda\, \frac{U(\mathbf{x})}{c^2} +O(\lambda ^2)\quad (\mathrm{no\ sum\ on\ }j).
\ee
This will indeed be the case if a post-Newtonian family of static bodies is envisaged either in the framework of GR in the harmonic gauge \cite{Fock64,Weinberg,FutaSchutz,A36}, or in the framework of a recent scalar theory \cite{A35}. In both cases, $U$ is {\it formally} identical to the Newtonian potential: it is defined in the coordinates $t,{\bf x}$ by
\be \label{U}
U(\mathbf{x}) \equiv G \int  \rho  (\mathbf{y})\ \dd ^3\mathbf{y}/\abs{\mathbf{x - y}},
\ee
where $\rho $ arises as the first coefficient in the expansion of the energy density $T^{00\,(\lambda )}$ in the coordinates $t,{\bf x}$, with ${\bf T}^{(\lambda )}$ the energy-momentum tensor in system S$_\lambda $:
\be \label{T00-expansion} 
T  ^{00(\lambda )}=\lambda \rho \left[1  + O(\lambda )\right].
\ee
Note that, therefore, the first  approximation of the energy density in system S$_\lambda $ (the ``Newtonian density") is $ \rho_\mathrm{N}^{(\lambda )}=\lambda \rho$. Hence, the real equivalent of the Newtonian potential is not $U$ (which does not depend on $\lambda $), but is instead, in system S$_\lambda $:
\be \label{U_N}
U_\mathrm{N}^{(\lambda )}(\mathbf{x}) \equiv \lambda U(\mathbf{x}) = G \int  \rho_\mathrm{N}^{(\lambda )}  (\mathbf{y})\ \dd ^3\mathbf{y}/\abs{\mathbf{x - y}}.\\
\ee

The {\it system of interest,} S, e.g. the Earth (assumed isolated), is supposed to correspond to a small finite value $\lambda _0 \ll 1$ of the parameter: $\mathrm{S}=\mathrm{S}_{\lambda _0}$, so that it makes sense to use the asymptotic expansions for that system. This means that the gravitational field in system S is indeed a weak field in the physical sense. 
\footnote{\ 
This is not automatical if one considers an a priori given, abstract family (S$_\lambda $), as is done in Ref. \cite{FutaSchutz}, because asymptotic expansions like (\ref{V-expansion})--(\ref{W-expansion}) remain true if one changes $\lambda $ to $\lambda '=\alpha \lambda $, with $\alpha $ a constant: in that case, the fact that $\lambda$ is small for some particular system $\mathrm{S}=\mathrm{S}_{\lambda _0}$ has no objective meaning. But in physical practice, one defines $\lambda _0$ in the very system of interest S, e.g. $\lambda _0 \equiv \mathrm{Sup}_{{\bf x}\in \mathsf{R}^3} (1-V({\bf x})) $ for the metric (\ref{g-Obukhov}). One must check that it is indeed negligible w.r.t. the unity: {\it this} means really that the gravitational field in system S is a weak field in the physical sense. Then, the family $(\mathrm{S}_{\lambda})$ is {\it deduced} from the data of S \cite{A36,A35}. (This is done by deducing a family of initial data from the initial data corresponding to S.)
} 
Moreover, the classical velocity of the test particle is assumed to be of the same order in the small parameter as that of a test particle orbiting in the weak gravitational field considered, namely \cite{Fock64,Weinberg,FutaSchutz,A36,A35}
\be \label{PNvelocity}
{\bf u}^{(\lambda )}(t)=\mathrm{ord}(\sqrt{\lambda})={\bf u}_0(t)\sqrt{\lambda} + O(\lambda ^{3/2}).
\ee
This means physically (in system S) that the initial velocity of the particle (e.g., the initial velocity of the neutron flux) is at most of the order of orbital velocities (as is certainly the case in the experiments \cite{COW1975,Nesvizhevsky2002,Nesvizhevsky2003}), so that it makes sense to use the PNA---then, ${\bf u}$ (and also $\dot{{\bf u}}$) will remain of PN magnitude as the time goes. It follows that the classical energy of the particle in the gravitational field \cite{L&L} admits the expansion
\be \label{E-mc2-classical}
E^{(\lambda )}_\mathrm{classical}\equiv \left \{g^{(\lambda )}_{00}\left[m^2c^4+({\bf p}^{(\lambda )})^2 c^2  \right]\right \}^{1/2} =mc^2+O(\lambda).
\ee

\subsection{Application to the stationary energy levels}
The stationary energy levels of the particle, considered as a Dirac (quantum) particle, must have the same order of magnitude as the particle's classical energy. Hence, any solution $E^{(\lambda )}$ of the eigenvalue problem (\ref{eigenvalue-E})$_\lambda $ (i.e., for system S$_\lambda $), should have an expansion 
\be \label{E-mc2} 
E^{(\lambda )}= mc^2 + O(\lambda )
\ee
(assuming that one may thus follow as a function of $\lambda $ each of the different eigenvalues). This asymptotic expansion of an energy level {\it implies that it is positive} (as soon as $\lambda $ is small enough). More precisely than (\ref{E-mc2}), we shall assume a first-order expansion for $E^{(\lambda )}$, or rather equivalently for $\omega ^{(\lambda )}\equiv E^{(\lambda )}/\hbar$: 
\be \label{E-expansion} 
\frac{\omega ^{(\lambda )}}{c}=M + \omega _1 \lambda + O(\lambda^2 ),
\ee
as well as for the amplitude function $a^{(\lambda )}=(\varphi ^{(\lambda )},\chi ^{(\lambda )})$:
\be \label{a-expansion} 
\varphi  ^{(\lambda )}=\varphi  _0 + \varphi  _1 \lambda + O(\lambda^2 ),\qquad \chi   ^{(\lambda )}=\chi  _0 + \chi  _1 \lambda + O(\lambda^2 ).
\ee
Inserting (\ref{E-expansion}) and (\ref{a-expansion}) into the explicit eigenvalue equations (\ref{eigenvalue-E-phi}) and (\ref{eigenvalue-E-chi}), using the expansions (\ref{V-expansion}) and (\ref{W-expansion}) of $V$ and $W$, and identifying powers of $\lambda $, yields, at the order zero,
\be \label{expanded-eigenvalue-0-1}
-i\sigma ^j \chi _{0,j} =0, 
\ee
\be \label{expanded-eigenvalue-0-2}
-i\sigma ^j \varphi  _{0,j} =2M\chi _0,
\ee
and at the order one (with $\tilde{U}\equiv U/c^2$):
\be \label{expanded-eigenvalue-1-1}
-i\sigma ^j \left(\chi _{1,j}-2\tilde{U} \chi _{0,j} \right) - \frac{i}{2}\sigma ^j \tilde{U} _{,j} \chi _0=\left(\omega _1+M\tilde{U}\right)\varphi _0,
\ee
\be \label{expanded-eigenvalue-1-2}
-i\sigma ^j \left(\varphi  _{1,j}-2\tilde{U} \varphi  _{0,j} \right) - \frac{i}{2}\sigma ^j \tilde{U} _{,j} \varphi  _0=2M\chi _1+\left(\omega _1-M\tilde{U}\right)\chi  _0.
\ee
Using the well-known property of the Pauli matrices:
\be \label{Pauli-product}
\sigma ^j \sigma ^k= \left \{\begin{array}{ll} i \varepsilon _{jkl}\sigma ^l & \mathrm{if}\ j\ne k\\
{\bf 1}_2 & \mathrm{if}\ j= k, \end{array} \right.
\ee
we note first that (\ref{expanded-eigenvalue-0-1}) and (\ref{expanded-eigenvalue-0-2}) imply
\be \label{Delta-phi_0=0}
\Delta \varphi _0=0.
\ee
Then we eliminate $\chi _0$ and $\chi _1$ from (\ref{expanded-eigenvalue-1-1}) by using (\ref{expanded-eigenvalue-0-2}) and (\ref{expanded-eigenvalue-1-2}), and we get with the help of (\ref{expanded-eigenvalue-0-1}), (\ref{Pauli-product}) and (\ref{Delta-phi_0=0}):
\be \label{expanded-eigenvalue-phi}
\frac{1}{2M}\left[ -\Delta \varphi _1 + \frac{3i}{2}\varepsilon _{jkl}\tilde{U} _{,j}\sigma ^l.\varphi _{0,k}+ \frac{1}{2}\tilde{U} _{,j} \varphi _{0,j} - \frac{1}{2}\varphi _0 \Delta \tilde{U} \right]
=\left(\omega _1+M\tilde{U}\right)\varphi   _0.
\ee
In Eqs. (\ref{Delta-phi_0=0}) and (\ref{expanded-eigenvalue-phi}), we have four independent scalar equations for the four scalar unknowns contained in the zero-order and first-order coefficients $\varphi _0$ and $\varphi _1$ of the expansion of the 2-component field $\varphi $. This makes the 1PN eigenvalue problem determinate. However, we may use Eq. (\ref{Delta-phi_0=0}) so as to rewrite (\ref{expanded-eigenvalue-phi}) as a single approximate equation for the 1PN field (which is the order-one approximation of $\varphi ^{(\lambda )}$)
\be \label{phi_(1)}
\varphi^{(\lambda )} _{(1)}\equiv \varphi _0 +\lambda \varphi _1.
\ee
Owing to (\ref{Delta-phi_0=0}) and (\ref{phi_(1)}), we have indeed in Eq. (\ref{expanded-eigenvalue-phi}):
\be
 \Delta \varphi _1 = \frac{1}{\lambda}\Delta \varphi^{(\lambda )}_{(1)}.
\ee
In the same way, it follows from (\ref{E-expansion}) and (\ref{phi_(1)}) that
\be
\omega _1 \, \varphi   _0 = \frac{1}{\lambda} \left(\frac{\omega^{(\lambda )} }{c}-M\right)\varphi^{(\lambda )}_{(1)}+O(\lambda ).
\ee
Thus, multiplying (\ref{expanded-eigenvalue-phi}) by $\lambda \hbar c$, we get [omitting the superscript $^{(\lambda )}$ and the subscript $_{(1)}$, i.e., defining $\varphi \equiv \varphi ^{(\lambda )}_{(1)}$, and also $U_\mathrm{N}\equiv U_\mathrm{N}^{(\lambda )}\equiv \lambda U$, Eq. (\ref{U_N})]:
\begin{eqnarray} \label{expanded-eigenvalue-phi_(1)}
-\frac{\hbar^2}{2m}\Delta \varphi -mU_\mathrm{N}\varphi + \frac{\hbar^2}{2mc^2}\left[\frac{3}{2}i\varepsilon _{jkl} U_{\mathrm{N} ,j} \sigma ^l  \varphi_{,k} +
\frac{1}{2} U_{\mathrm{N} ,j} \varphi_{,j}  -\frac{1}{2}\varphi\, \Delta U_\mathrm{N} \right ] & &\nonumber\\= (E-mc^2)\varphi +O(\lambda ^{2}).& & 
\end{eqnarray}
This equation shows explicitly the additional terms with respect to the stationary non-relativistic Schr\"odinger equation in the Newtonian gravity field,
\be \label{Schroedinger}
-\frac{\hbar^2}{2m}\Delta \varphi -mU_\mathrm{N}\varphi = E_\mathrm{nr} \varphi.
\ee
(The subscript ``nr" stands for ``non-relativistic.") It is interesting to note that, in Eq. (\ref{expanded-eigenvalue-phi_(1)}),  all terms are of the same order in the small parameter $\lambda $. [This is a priori obvious since (\ref{expanded-eigenvalue-phi_(1)}) is a mere rewriting of the exact equation (\ref{expanded-eigenvalue-phi}), the latter involving only {\it expansion coefficients} like $\varphi  _0,\  \varphi  _1$ and $U$, which are, of course, of order zero in the small parameter.] Namely, all terms are order $\lambda $.
\footnote{\ 
On the r.h.s. of (\ref{expanded-eigenvalue-phi_(1)}), we consider $(E-mc^2)\varphi$ as  one term: it is of order one in $\lambda $, but it is the difference between two terms of order zero in $\lambda $.
} 
This result means that, from the point of view of the asymptotic PN scheme, the corrections to the non-relativistic Schr\"odinger equation (\ref{Schroedinger})---e.g., the corrections to the energy levels--- do not need to be small with respect to the corresponding quantities as they are found using Eq. (\ref{Schroedinger}). However, an asymptotic estimate is not a numerical one: we must now numerically investigate the differences in the energy spectrum. 

\subsection{Estimates for ultra-cold neutrons in the Earth's gravity}
The mass of the neutron is $m\simeq \frac{1}{6}\times 10^{-26}\,$kg ($mc^2\simeq 939.57\,$MeV $\simeq 1.50\times 10^{-10}\,$J). The Earth's Newtonian potential is, on Earth, $U_{\mathrm{N}}\simeq \frac{GM_\oplus }{r_\oplus }\simeq \frac{6.67\times 10^{-11}\times 6\times 10^{24}}{6.37\times 10^6}\simeq 6.3\times 10^7\, \mathrm{m}^2/\mathrm{s}^2$, whence $mU_{\mathrm{N}}\simeq 10^{-19}\,$J. We have $\abs {\nabla U_\mathrm{N} } = g\simeq 9.81\,\mathrm{m}.\mathrm{s}^{-2}$. Assuming perturbatively for $\varphi $ the functions found from the non-relativistic Schr\"odinger equation for the first energy levels of ultra-cold neutrons in the Earth's gravity \cite{Nesvizhevsky2003}, we have $\abs {\nabla \varphi }\simeq 10^5 \,\mathrm{m}^{-1}$. Now the relativistic corrections involve the minute coefficient 
\be \label{alpha}
\xi  \equiv \frac{\hbar^2 }{2mc^2}\simeq \frac{(1.054\times 10^{-34})^2\times 3\times 10^{26}}{9\times 10^{16}}\simeq 3.7 \times 10^{-59}\,\mathrm{kg}.\mathrm{m}^2.
\ee
Therefore, the two first correction terms in the square bracket in Eq. (\ref{expanded-eigenvalue-phi_(1)}), which are of the order of $\xi g \abs {\nabla \varphi }\simeq 3\times 10^{-53}$J, are utterly negligible with respect to $mU_{\mathrm{N}}\varphi \simeq 10^{-19}\,$J. It is even more so for the third, last correction term in (\ref{expanded-eigenvalue-phi_(1)}), $-\frac{\xi }{2} \varphi \Delta U_{\mathrm{N}}$ with $-\Delta U_{\mathrm{N}}=4\pi G \rho_\mathrm{N} \simeq 8\times 10^{-7}\, \mathrm{s}^{-2}$ (with $\rho_\mathrm{N} \simeq 10^3\,$kg/m$^3$).\\

However, one should account for the fact that the energy levels are defined only up to a constant: in Eq. (\ref{expanded-eigenvalue-phi_(1)}), one may change simultaneously $ U_\mathrm{N}$ to $U'_\mathrm{N}=U_\mathrm{N}+C$ and $E$ to $E'=E-mC$. [The same is true in Eq. (\ref{Schroedinger}).] It follows that, in the experiments with neutrons passing through a horizontal slit \cite{Nesvizhevsky2003}, which are very localized as compared with the Earth's radius, we may approximate the Newtonian potential as
\be
U_{\mathrm{N}} \simeq -gz,
\ee
with $z$ the vertical coordinate, counted upwards from the lower side of the slit, and taking values up to a few $ 10^{-5} \,$m. This gives now $-mU_{\mathrm{N}} \simeq 10^{-30}\,$J. That has the same magnitude as the non-relativistic energies $E_\mathrm{nr} \simeq E-mc^2$ \cite{Nesvizhevsky2003}. It still exceeds the relativistic corrections by some 23 orders of magnitude.

\section{Conclusion}\label{Conclusion}

In this work, we derived the post-Newtonian equation for the stationary energy levels of a slow Dirac(-Fock-Weyl) (DFW) particle in a weak static gravitational field, Eq. (\ref{expanded-eigenvalue-phi_(1)}). To our knowledge, this equation was not derived before, although weak-field expressions for the DFW Hamiltonian have been proposed \cite{deOliveiraTiomno1962,VarjuRyder1998,Obukhov2001,SilenkoTeryaev2005}. In a first step, we showed that there is just one natural coordinate-independent scalar product relevant to stationary wave functions in a static metric, given by Eqs. (\ref{Hermitian-general})--(\ref{invariant-V}). In the case of a diagonal tetrad, this scalar product coincides with the scalar product (\ref{Hermitian-gamma}) considered by Leclerc \cite{Leclerc2006a}, and which arises from the conservation law of the DFW current. We checked directly that, at least in the case where the metric can be set in the ``space-isotropic" form (\ref{g-Obukhov}) assumed by Obukhov \cite{Obukhov2001}, the Hamiltonian (\ref{Hamilton-DFW}) of the DFW equation turns out to be Hermitian for that scalar product. Instead, de Oliveira \& Tiomno \cite{deOliveiraTiomno1962}, Varj\'u \& Ryder \cite{VarjuRyder1998} and Obukhov \cite{Obukhov2001} used the non-unitary transformation (\ref{Non-unitary-transform}). That transformation gets the scalar product (\ref{Hermitian-gamma}) to the ``flat product" (\ref{Hermitian-coordinate}), with which the hermiticity of the modified Hamiltonian, equivalent to that of the starting Hamiltonian for the product (\ref{Hermitian-gamma}), is easily checkable.\\

In a second step, we used the ``asymptotic" scheme of post-Newtonian approximation (PNA) \cite{FutaSchutz,A36,A35} to determine the approximate equation that governs the stationary energy levels, in order to be able to compare that equation with the non-relativistic Schr\"odinger equation in the gravity potential. To use that scheme, we assumed that the gravitational field is weak and that the classical velocity of the Dirac particle is small, both in an asymptotic sense, i.e., for $\lambda \rightarrow 0$---considering a family (S$_\lambda $) of systems. It makes sense to apply the results to the system of interest (e.g., a flux of ultra-cold neutrons in a laboratory on the Earth), insofar as the relevant value $\lambda _0$ of the parameter for that system is small (as is indeed the case in the example mentioned). In our final equation (\ref{expanded-eigenvalue-phi_(1)}), the new terms, as compared with the non-relativistic Schr\"odinger equation (\ref{Schroedinger}), are of the same order in the small parameter as the terms which are already there in Eq. (\ref{Schroedinger})---namely, they are order one. This is a surprising result if one compares it with the situation for classical particles, in which situation the first relativistic corrections are one order higher in the small parameter than the Newtonian terms. However, this result does not mean that the corrections have the same numerical magnitude as the ``main" terms. And indeed, we found that, for ultra-cold neutrons in the Earth's gravity field, the corrections of the DFW equation to the energy levels are hopelessly negligible with respect to the energy levels as found with the non-relativistic Schr\"odinger equation.  \\

We note that, in their works on the nonrelativistic limit of the DFW equation, de Oliveira \& Tiomno \cite{deOliveiraTiomno1962}, Varj\'u \& Ryder \cite{VarjuRyder1998}, Obukhov \cite{Obukhov2001}, as well as Silenko \& Teryaev \cite{SilenkoTeryaev2005}, used Foldy-Wouthuysen \cite{BjorkenDrell1964,FoldyWouthuysen1950} transformations, or at least (according to the discussion in Ref. \cite{SilenkoTeryaev2005}), transformations of the Foldy-Wouthuysen type. In contrast with these authors, we consider explicitly the {\it stationary energy levels}, and this {\it in the post-Newtonian approximation}. These two points, taken together, allow us to state Eq. (\ref{E-mc2}) and to use it, thus automatically selecting the positive-energy solutions. For this reason, we did not need to have recourse to a Foldy-Wouthuysen transformation. \\

{\bf Acknowledgement.}  I am grateful to F. Selleri, F. Romano, to the late C. De Marzo and to his nice team, for their warm hospitality in Bari. Also, thanks are due to Franco Selleri and to Konstantin Protasov, who urged me to evaluate the magnitude of the corrections to the energy levels. Finally, the referee pointed out the very relevant work \cite{Leclerc2006a} and asked questions about the change of units used in the previous version \cite{A38v1}, which led me to realize that this change is unnecessary in the present case---thus simplifying the redaction.


\begin{thebibliography}{9}
\small
\bibitem{KieferWeber2005}
C. Kiefer and C. Weber, {\it Ann. der Phys.} {\bf 14}, 253 (2005). [gr-qc/0408010]

\bibitem{COW1975}
R. Colella, A. W. Overhauser and S. A. Werner, {\it Phys. Rev.
Lett.} {\bf 34}, 1472 (1975).

\bibitem{RiehleBorde1991}
F. Riehle, Th. Kisters, A. Witte, J. Helmcke and Ch. J. Bord\'e, {\it Phys.
Rev. Lett.} {\bf 67}, 177 (1991).

\bibitem{KasevichChu1991}
M. Kasevich and S. Chu, {\it Phys. Rev. Lett.} {\bf 67}, 181 (1991).

\bibitem{Nesvizhevsky2002} 
V. V. Nesvizhevsky {\it et al.}, {\it Nature} {\bf 415}, 297 (2002).

\bibitem{Nesvizhevsky2003} 
V. V. Nesvizhevsky {\it et al.}, {\it Phys. Rev. D} {\bf 67}, 102002 (2003).  [hep-ph/0306198]

\bibitem{Nesvizhevsky2005} 
V. V. Nesvizhevsky {\it et al.}, {\it Eur. Phys. J. C} {\bf 40},  479 (2005). [hep-ph/0502081]

\bibitem{OverhauserColella1974}
A. W. Overhauser and R. Colella, {\it Phys. Rev. Lett.} {\bf 33}, 1237 (1974).

\bibitem{LuschikovFrank1978}
V. I. Luschikov and A. I. Frank, {\it JETP Lett.} {\bf 28}, 559 (1978).

\bibitem{WestphalAbele2006}
A. Westphal, H. Abele, S. Bae\ss{}ler, V. V. Nesvizhevsky, A. K. Petukhov, K. V. Protasov and A. Yu. Voronin, hep-ph/0602093 (2006).

\bibitem{BirrellDavies}
N. D. Birrell, P. C. W. Davies, {\it Quantum Fields in Curved Space} (Cambridge University Press, Cambridge, 1982), Section 3.2.

\bibitem{Fock1929b}
V. A. Fock, {\it Z. Phys.} {\bf 57}, 261 (1929).

\bibitem{Weyl1929b}
H. Weyl, {\it Z. Phys.} {\bf 56}, 330 (1929).

\bibitem{A37}
M. Arminjon, {\it Found. Phys. Lett.} {\bf 19}, 225 (2006). [gr-qc/0512046]

\bibitem{deOliveiraTiomno1962}
C. G. de Oliveira and J. Tiomno, {\it Nuovo Cim.} {\bf 24}, 672 (1962).

\bibitem{VarjuRyder1998}
K. Varj\'u and L. H. Ryder, {\it Phys. Lett. A} {\bf 250}, 263 (1998).

\bibitem{Obukhov2001}
Yu. N. Obukhov, {\it Phys. Rev. Lett.} {\bf 86}, 192 (2001). [gr-qc/0012102]

\bibitem{SilenkoTeryaev2005}
A. J. Silenko and O. V. Teryaev, {\it Phys. Rev. D} {\bf 71}, 064016 (2005). [gr-qc/0407015]

\bibitem{A38v1}
M. Arminjon, first version of the present work, gr-qc/0606036v1.

\bibitem{Leclerc2006a}
M. Leclerc, {\it Class. Quant. Grav.} {\bf 23}, 4013 (2006). [gr-qc/0511060]

\bibitem{BrillWheeler1957+Corr}
D. R. Brill and J. A. Wheeler, {\it Rev. Modern Phys.} {\bf 29}, 465 (1957). Erratum: {\it Rev. Modern Phys.} {\bf 33}, 623 (1961).

\bibitem{VillalbaGreiner2001}
V. M. Villalba and W. Greiner, {\it Phys. Rev. D} {\bf 65}, 025007 (2001). [gr-qc/0112006]

\bibitem{BjorkenDrell1964}
J. D. Bjorken and S. D. Drell, {\it Relativistic Quantum Mechanics} (McGraw-Hill, New York etc., 1964).

\bibitem{A16}
M. Arminjon, {\it Arch. Mech.} {\bf 48}, 551 (1996). (Online at \verb+geo.hmg.inpg.fr/arminjon/pub_list.html#A16+ .)

\bibitem{L&L} 
L. Landau, E. Lifchitz, {\it Th\'eorie des Champs} ($4^{\mathrm{th}}$ French edn., Mir, Moscow, 1989). 

\bibitem{B15} 
M. Arminjon, in {\it Sixth Int. Conf. Physical Interpretations of Relativity Theory, Proceedings}, M.C. Duffy, edr. (British Soc. Philos. Sci. and University of Sunderland, Sunderland, 1998), pp. 1-17. [gr-qc/0203104]

\bibitem{DieudonneTome2}
J. Dieudonn\'e, {\it El\'ements d'Analyse, tome 2} (Hermann, Paris, 1969), p. 357.

\bibitem{AudretschSchaefer1978}
J. Audretsch and G. Sch\"afer, {\it Gen. Rel. Grav.} {\bf 9}, 243 (1978).

\bibitem{KieferSingh1991}
C. Kiefer and T. P. Singh, {\it Phys. Rev. D} {\bf 44}, 1067 (1991).

\bibitem{Laemmerzahl1995}
C. L\"ammerzahl, {\it Phys. Lett. A} {\bf 203}, 12 (1995).

\bibitem{FutaSchutz}
T. Futamase and B. F. Schutz, {\it Phys. Rev. D} {\bf 28},
2363 (1983).

\bibitem{A36} 
M. Arminjon, {\it Phys. Rev. D} {\bf 72}, 084002 (2005). [gr-qc/0504016]

\bibitem{A35}
M. Arminjon, {\it Brazil. J. Phys.} {\bf 36}, 177 (2006). [gr-qc/0412085]

\bibitem{Fock64}
V. A. Fock, {\it The Theory of Space, Time and Gravitation} (2nd
English edn., Pergamon, Oxford, 1964). (First Russian edition 1955.)

\bibitem{Weinberg}
S. Weinberg, {\it Gravitation and Cosmology} (J. Wiley \& Sons, New York, 1972).

\bibitem{FoldyWouthuysen1950}
L. L. Foldy and S. A. Wouthuysen, {\it Phys. Rev.} {\bf 78}, 29 (1950).

\end{thebibliography}
\end{document}